\documentclass[12pt]{article}
\usepackage{amsfonts}
\begin{document}
\title{The Shr\"odinger operator on graphs and topology}
\author{S.\,P.\,Novikov}
\date{}
\maketitle

We shall define a Schr\"odinger operator for a one-dimensional simplicial
complex (a graph)\footnote{We began the discussion of this
problem together with A.\,P.\,Veselov}
$\Gamma$ without ends (that is, at least two and only finitely
many edges meet at any vertex),
which acts on functions of vertices $T$ or edges $R$:
$$(L\psi)_T=\sum\limits_{T'}b_{T:T'}\psi_{T'}\eqno{\rm(V)}$$
for $L_0=\partial\partial^*$, $b_{T:T'}=1$, $b_{T:T'}=-m_T$,
$$(L\psi)_R=\sum\limits_{R'}d_{R:R'}\psi_{R'}\eqno{\rm(E)}$$
for $L_0=\partial^*\partial$, $d_{R:R'}=1$, $d_{R:R}=-2$.

Here the coefficients are real, symmetric, and non-zero only for
nearest neighbours
$$T'\cup T=\partial R,\qquad R\cap R'=T.$$

The coefficients $b_{T:T}=W_T$ and $d_{R:R}=W_R$ are
called the potential, $m_T$ is the number of edges at a vertex,
and $\partial$ is the boundary operator.

\medskip
\noindent{\bf Definition 1.}
The {\it Wronskian\/} of a pair of solutions of $L\psi_i=\lambda
\psi_i$, $i=1,2$, is the skew-symmetric bilinear expression
$$W_{\mathbf R}=b_{T:T'}(\psi_{1T}\psi_{2T'}-\psi_{2T}\psi_{1T'}),\quad
T'\ne T,\eqno{\rm(V)}$$
$$W_{\mathbf R}=\sum\limits_{R'\cap R=T}d_{R:R'}(\psi_{1R}
\psi_{2R'}-\psi_{2R}\psi_{1R'}),\qquad R'\ne R.\eqno{\rm(E)}$$
Here ${\mathbf R}=(TT')$ is an oriented edge, $W_{TT'}=-W_{T'T}$.

\medskip
\noindent{\bf Theorem 1.} {\it The Wronskian of a pair of solutions
is well defined as a 1-chain on $\Gamma$, whose boundary
is equal to zero: $\partial W=0$.}

\medskip
The proof follows in both cases by considering the quantity $(L\psi_1)
\psi_2-(L\psi_2)\psi_1$.

We consider a graph $\Gamma$ having $k$ `tails' $(z_1,\dots,z_k)$,
$k\ge1$: the tails $z_j$ are half-lines with edges
$R_{jn}$, $n\ge1$, and vertices $T_{j,n-1}$,
$\partial R_{jn}=T_{jn}-T_{j,n-1}$. Suppose that the Schr\"odinger
operator is `finitary', that is, for $n>n_0$ and all $j$ we
have $L=L_0+2$ in both cases.

The equation $L\psi=\lambda\psi$ has solutions in $z_j$ as
$n\rightarrow\infty$ of the form $\psi_{jn}^\pm=a_\pm^n$,
$a_\pm=\frac12(\lambda\pm\sqrt{\lambda^2-4})$. A basis of real
solutions for $\lambda\in\mathbb R$ is:
$$C_{jn}=(a_-\psi_{jn}^++a_+\psi_{jn}^-)(a_++a_-)^{-1},\qquad
S_{jn}=(\psi_{jn}^+-\psi_{jn}^-)(a_+-a_-)^{-1}.$$

These solutions are defined only on the tails. We introduce a `phase space'
$\mathbb R^{2k}$ with basis $(C_1,S_1,\dots,C_k,S_k)$ and a skew-scalar
product
$$\langle C_i,C_j\rangle=\langle S_i,S_j\rangle=0,\qquad
\langle C_i,S_j\rangle=\delta_{ij}.$$

The operator $L$ determines the subspace $\Lambda_\lambda^\infty\subset
\mathbb R^{2k}$ of vectors $\psi_\infty\in\Lambda_\lambda^\infty$,
which can be extended to the whole graph $\Gamma$ as the solution
$$L\psi=\lambda\psi,\qquad
\psi=\psi_0=(\psi_{1\infty},\dots,\psi_{k\infty}),\quad
\psi_{j\infty}=\alpha_jC_j+\beta_jS_j$$
in the tail $z_j$ for $n>n_0$.

\medskip
\noindent{\bf Theorem 2.} \footnote{In 1971 I.\,M.\,Gelfand
told the author about an interesting idea on the relation
between self-adjoint extensions of symmetric operators and Lagrangian
planes. This was Gel'fand's reaction to the author's paper
\cite{2}, built round the relation between Hamiltonian formalism
and differential topology.}
{\it The subspace $\Lambda_\lambda^\infty$ is Lagrangian,
that is, the scalar product on it is equal to zero.}

\medskip
The proof follows from Theorem 1. In fact, for any pair of solutions
$\psi_1$, $\psi_2$ on $\Gamma$ we have $W=\sum_{j=1}^k
\kappa_jz_j+\mbox{ (finite)}$, where the $z_j$ are the tails.
Only the differences $z_i-z_j$ can be extended to cycles
on $\Gamma$ modulo $\infty$. Therefore we have $W=\sum_{l\ge2}^k
\mu_l(z_1-z_l)+\mbox{ (a finite cycle)}=\sum_{j=1}^k
\kappa_jz_j+\mbox{ (a finite chain)}$. Hence the theorem follows:
$$\sum\limits_{j=1}^k\kappa_j=0=\langle\psi_{1\infty},\psi_{2\infty}
\rangle.$$

We represent the graph $\Gamma$ in the form $\Gamma=\Gamma'\cup
K_1\cup\cdots\cup K_s$, where the $K_l$ are trees growing
from the vertices (`nests') $T_l\in\Gamma'$, with $\Gamma'$
a finite graph without ends (`the base'), $s\le k$.

\medskip
\noindent{\bf Theorem 3.}
{\it The dimension of the spaces of solutions of $L\psi=\lambda\psi$
on $\Gamma$ is always at least $k$. It is strictly greater
than $k$ if and only if the Schr\"odinger operator $L'$,
restricted to the base $\Gamma'$ has
a `singular' eigenvalue $\lambda_p'$, where $\psi_P'(t_l)=0$ at
all nests $l=1,\dots,s$. In this case and anly in this case
the operator that associates the vector $\psi_\infty\in
\Lambda_\lambda^\infty$ with the solution of $L\psi=\lambda\psi$
has a non-trivial kernel.}

\medskip
The spectrum of the operator $L$ in the Hilbert space
${\cal L}_2^\varepsilon(\Gamma)=H_\varepsilon$, $\varepsilon=0,1$ (vertices
and edges) can be divided into a continuous part $|\lambda|\le2$
(scattering zone) and a discrete spectrum in the zones
$\lambda<-2$, $\lambda>2$. Points of the discrete spectrum inside
the zone $|\lambda|\le2$ are possible only if localized on
$\Gamma'$, where $\lambda=\lambda'$ is a singular eigenvalue.
The algebraic properties of the scattering zone ({\bf unitarity}) are
entirely determined by Theorem 2.
For $|\lambda|>2$ we have $a_+a_-=1$, $a_\pm\in\mathbb R$.
We consider a Lagrangian plane $\Lambda_\lambda^-\subset\mathbb
R^{2k}$ with  basis $\Psi_j^-$, $j=1,\dots,k$, decreasing at
infinity.

\medskip
\noindent{\bf Proposition 1.}
{\it A point $\lambda\in\mathbb R$, is a point of the discrete
spectrum for $L$ in $H_\varepsilon$ $(\varepsilon=0,1)$
if $\Lambda_\lambda^\infty\cap\Lambda_\lambda^-$ is non-empty or
$\lambda=\lambda_p'$ is a singular eigenvalue on the base $\Gamma'$
The Morse indices of eigenvalues for $\lambda\in[2,\infty]$
and $\lambda\in[-\infty,-2]$ are well-defined and equal to the
algebraic numbers of the normal discrete eigenvalues.
 There is necessarily a discrete
spectrum if: $\max_T\left(\sum b_{T:T'}^2+W_T^2\right)>4$,
$\max_R\left(\sum d_{R:R'}^2+W_R^2\right)>4$, or
$\max_q|\lambda_q'|>4$ for the spectrum of base $\Gamma'$.}

\end{document}